\documentclass[global,twocolumn]{svjour}
\usepackage{latexsym}
\usepackage{graphicx}

\journalname{myjournal}
\newcommand{\beq}{\begin{equation}}
\newcommand{\eeq}{\end{equation}}
\newcommand{\beqa}{\begin{eqnarray}}
\newcommand{\eeqa}{\end{eqnarray}}
\newcommand{\nnb}{\nonumber}
\begin{document}
\title{Probing vacuum birefringence by 
       phase-contrast Fourier imaging under fields
       of high-intensity lasers}

\author{K.~Homma\inst{1}${}^,$\inst{2}, D.~Habs\inst{2} and T.~Tajima\inst{2}}

\offprints{Kensuke Homma}

\institute{
    $^1$ Graduate School of Science, Hiroshima University, Kagamiyama, 
         Higashi-Hiroshima 739-8526, Japan\\
    $^2$ Fakult\"at f\"ur Physik, Ludwig-Maximilians 
         Universit\"at M\"unchen, D-85748 Garching, Germany
}
\date{Received: date / Revised version: date}


\maketitle
\begin{abstract}
In vacuum high-intensity lasers can cause photon-photon interaction
via the process of virtual vacuum polarization
which may be measured by the phase velocity 
shift of photons across intense fields.
In the optical frequency domain, the photon-photon interaction 
is polarization-mediated described by the Euler-Heisenberg effective 
action. This theory predicts the vacuum birefringence or polarization 
dependence of the phase velocity shift 
arising from nonlinear properties in quantum electrodynamics (QED). 
We suggest a method to measure the vacuum birefringence under intense
optical laser fields based on the absolute phase velocity shift
by phase-contrast Fourier imaging. The method may serve for
observing effects even beyond the QED vacuum polarization.
\end{abstract}

\sloppy
\section{Introduction}\label{sec1}
To observe nonlinear responses of matter, the pump-probe technique
is widely used: Matter is first excited by an intense laser pulse and 
then probed by a delayed weaker laser pulse.
When the vacuum is considered as a part of matter,
the most natural approach to probe it is, hence, the pump-probe technique.
Maxwell's equations in vacuum, however, allow only for linear superpositions
of laser fields. 
In quantum mechanics,
a photon can be resolved into a pair of virtual fermions over a short time 
via the uncertainty principle in the higher frequency domain even below 
the fermion mass scale.
The loop of the virtual pair provides a coupling to photons, resulting in a
photon-photon interaction.
In the optical frequency domain, the
electron-positron loop and possibly the lightest quark-antiquark loop are 
expected to give rise to the photon-photon interaction
with the mass scale of the electron being $0.5$~MeV/$c^2$ and
of the lightest quark ranging from $\sim 1 - 100$~MeV/$c^2$, respectively.
Below the electron mass scale, 
there is no known mass scale relevant for photon-photon
interactions in the standard model of particle physics.
In this paper we focus on the photon-photon interaction
in the optical laser frequency range based on quantum electrodynamics (QED).


In the low-frequency collision $\hbar\omega \ll m_e c^2$,
it is sufficient to describe
the photon-photon interaction by the effective one-loop Lagrangian
\cite{EH,Weiscop,Schwinger}
%
\beqa\label{eq_EHL}
    L_{1-loop} =
    \frac{1}{360}\frac{\alpha^2}{m_e^4}
    [4(F_{\mu\nu}F^{\mu\nu})^2+7(F_{\mu\nu}\tilde{F}^{\mu\nu})^2],
\eeqa
where 
$\alpha=\frac{e^2}{\hbar c}$ is the fine structure constant,
$m_e$ is the electron mass,
$F_{\mu\nu} = \partial A_{\mu} / \partial x^{\nu} -
                    \partial A_{\nu} / \partial x^{\mu}$ is
the antisymmetric field strength tensor and its dual tensor
$\tilde{F}^{\mu\nu} = 1/2 \varepsilon^{\mu\nu\iota\rho} F_{\iota\rho}$
with the Levi-Civita symbol $\varepsilon^{\mu\nu\iota\rho}$.
%
%

\begin{figure}
\includegraphics[width=1.0\linewidth]{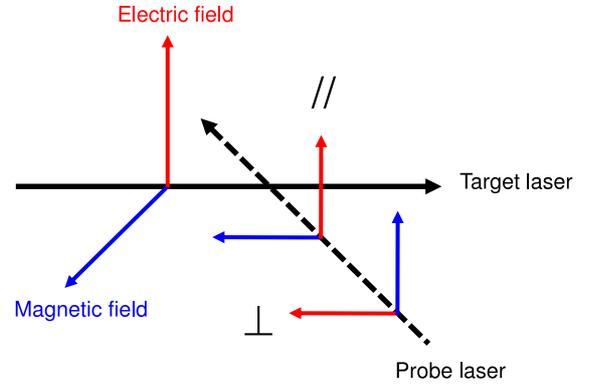}
\caption{
Polarization directions of the target and probe lasers.
}
\label{Fig1}
\end{figure}
Based on this Lagrangian, the dispersion relation for photons in vacuum
is expected to be modified by intense electromagnetic fields.
This effect under a constant electromagnetic field was first discussed
by Toll~\cite{Toll}. At optical frequencies, we may approximate
the time-varying electromagnetic field as a constant field,
because the relevant time scale for the creation
of virtual electron-positron pairs
is much shorter than that of the inverse of optical frequencies.
We can discuss the dispersion relation and
the birefringent nature via measurements of
the refractive index, {\it i.e.}, the inverse of the phase velocity
as illustrated in Fig.\ref{Fig1},
where a linearly polarized probe laser beam
crosses a linearly polarized target laser beam. The measurements of
the phase velocity shift when the
electric fields of both lasers are either parallel or normal to each other
are specified with subscriptions $\parallel$ or $\perp$, respectively.
The theoretical derivation of these quantities
in the linearly polarized electromagnetic field of the target
(the so-called crossed-field configuration, 
where the electric field $\hat{E}$ and
magnetic field $\hat{B}$ are normal at the same strength)
was originally studied in~\cite{Narozhnyi,Ritus} and further derived
from the generalized prescription based on the polarization tensor,
applicable to arbitrary external fields, in~\cite{Dittrich-Gies}.
This results in
%
\begin{eqnarray}\label{eq_phsv}
      v_{\parallel}/c = 1 - \frac{8}{45}\alpha^2 \frac{\rho_t}{\rho_c},
\nonumber \\
      v_{\perp}/c    = 1 - \frac{14}{45}\alpha^2 \frac{\rho_t}{\rho_c},
\end{eqnarray}
%
where $v_{\parallel}/c$ and $v_{\perp}/c$ are the phase velocities
when the combination of linear polarizations of the probe and target lasers is
parallel and normal, respectively. 
The quantity $\rho_c \equiv m^4_e c^5 /\hbar^3 
\sim 1.42\times10^{6}$J/$\mu$m${}^3$
is the Compton energy density of an electron
and $\rho_t$ is defined as $z_k/k^2$ where
$k$ is the wave number of the probe
electromagnetic field with the unit vector of $\hat{k}$.
The Lorentz-invariant quantity $z_k$ is defined as
%
\begin{eqnarray}\label{eq_zk}
    z_k = (k_{\alpha}F^{\alpha\kappa})(k_{\beta}F^{\beta}_{\kappa}),
\end{eqnarray}
%
and the relation to the energy density $\epsilon^2$ in the crossed field
condition is
%
\begin{eqnarray}\label{eq_zk_epsilon}
   \frac{z_k}{k^2} = \epsilon^2 (1+(\hat{k}\cdot\hat{n}))^2,
\end{eqnarray}
%
with $\epsilon = E = cB$ and $\hat{n} = \hat{B} \times \hat{E}$ with $\hat{}$
indicating the unit vector.
Thus the second terms in Eq.~(\ref{eq_phsv}) show that the deviation of
the phase velocities of light $v_{\parallel}$ and $v_{\perp}$
are proportional to the field energy density normalized to the
Compton energy density of an electron.
The shift of the refractive index from that of the normal vacuum is
on the order of $10^{-11}$ for the energy density
$\epsilon^2$ of 1~J/$\mu$m${}^3$ corresponding to the power density
of a high-power laser beam focused to $10^{22}$W/cm${}^2$ at its waist. 
The refractive medium exhibits a polarization
dependence, {\it i.e.}, it shows birefringence. 
The difference in $v_{\parallel}$
and $v_{\perp}$ in Eq.~(\ref{eq_phsv}) results from
the first and second terms in the bracket
of the effective one-loop Lagrangian in Eq.~(\ref{eq_EHL}).

The dispersion relation and the birefringence under a constant
electromagnetic field in the UV limit ($\omega\rightarrow\infty$)
may be evaluated via the Kramers-Kronig
dispersion relation, as discussed in~\cite{Shore}.
The phase velocity in both UV and IR is expected to be 
subluminal ($v_{phase} < c$) under the influence of 
the QED field ~\cite{Shore,Dittrich-Gies}.
The UV limit of the phase velocity is supposed
to govern causality which should not exceed the velocity of light
in vacuum. Therefore, it can be a fundamental test of a variety
of effective field theories in the IR by testing whether the phase velocity 
in the UV limit, extrapolated from that of the IR, is superluminal
($v_{phase}(\infty) > c$) or not.
Thus far the dispersion relation from IR to UV is theoretically known only
in the QED field~\cite{Shore}. However, there is no data so far even
in the domain of IR frequencies. It is important, therefore, 
for experiments to quantitatively verify or disprove the QED prediction.
We note that the measurement of the refractive index in the domain of
higher frequencies may be sensitive to the part of the anomalous dispersion
where the real part of the refractive index rises
as discussed in ~\cite{Heinzl}, and, also, the measurement of the 
electron-positron pair creation~\cite{Dunne,Baier,Narozhny2,Schuetzhold}
in strong electromagnetic fields may be directly sensitive to
the absorptive or imaginary part.
The Kramers-Kronig relation connects the real and imaginary parts
of the forward scattering amplitude or the refractive index.
Therefore, the systematic measurements of real and imaginary parts over
a wide frequency range may provide a test ground of QED and
the Kramers-Kronig relation itself, when it is applied to the vacuum.

The key issue is how to detect the extremely small refractive index change,
resulting from the photon-photon interaction
between the target and probe lasers.
The conventional way in the X-ray frequency range is based on a measurement of
the ellipticity caused by the target field-induced birefringence with respect to
the linearly polarized incident probe photons~\cite{Heinzl,LaserDiffraction}.
Since nowadays high-precision X-ray polarimetery technique is 
available~\cite{X-rayPol},
we may reach the sensitivity to QED-induced birefringence, if
high-intensity lasers such as those attainable in ELI~\cite{ELI} are provided.
As explained above, the probe frequency dependence of the birefringence
is important to complete the QED-induced dispersion relation. 
Therefore, we need measurements in the optical frequency range as well.
The conventional ways in the range of optical frequencies that were
 performed~\cite{PVLAS} and proposed~\cite{LaserLaser} are again based 
on a measurement of the ellipsoid caused by the birefringence and 
a measurement of the rotation angle of a linearly polarized probe laser 
by  making it propagate for a long distance under the influence of a weak 
magnetic~\cite{PVLAS} or electromagnetic field~\cite{LaserLaser}.
This method has the advantage to enhance the phase shift
by a long optical path without introducing costly strong target
electromagnetic fields.
In the case of a constant magnetic field on the order of 1~T,
one encounters the limits of physical sensitivity
to the QED nonlinear effects. In the case of an electromagnetic field,
we may be sensitive to the QED-induced birefringence within a few days 
with a 1J CW laser according to the claim in \cite{LaserLaser}.
However, if one aims at the sensitivity even beyond QED-induced birefringence
as we discuss in section~\ref{sec3}, it is essential to introduce a
high-intensity pulse even beyond the capability of the ELI facility~\cite{ELI}. 
In such circumstances the storage of a high-intensity laser pulse 
in a cavity is limited by the damage threshold of the optical 
elements needed to store the target field over a long time.

On the other hand, if we could localize the field-induced refractive index
change by tightly focusing a high-intensity target laser pulse and 
measuring the spatially inhomogeneous phase effect of the vacuum on 
a pulse-by-pulse basis,
there will be no physical limit in increasing the intensity of the laser pulse
until the vacuum itself breaks down. In order to increase the shift of the
refractive index, corresponding to the inverse of the phase velocities in
Eq.~(\ref{eq_phsv}), {\it i.e.}, the intensity of the target laser pulse
as expected from Eq.~(\ref{eq_zk}) and Eq.~(\ref{eq_zk_epsilon}),
it is necessary to use a focused
laser pulse by confining the large laser energy into a small space-time volume.
This causes a locally varying refractive index along the
trajectory of the target laser pulse in vacuum. A variation of
the refractive index arises over the high-intensity
part and the remaining vacuum. If the probe laser penetrates into both parts
simultaneously, the corresponding phase contrast should be embedded in the
transverse profile of the same probe laser. Our suggestion is 
to directly measure the phase contrast and to determine the absolute 
value of the refractive
index change by controlling the combination of polarizations of the probe
and target laser pulses. This should result in the birefringence
as expected in Eq.~(\ref{eq_phsv}). The birefringence measurement based on 
the measurement of absolute phase velocities we suggest here
should be contrasted to any existing techniques to measure 
the ellipticity where only relative phase differences can be discussed.

In the following sections we introduce the basic concept of
the phase-contrast Fourier imaging by crossing target and probe lasers
and discuss a way to extract a physically induced phase in the presence of 
a phase background. We then discuss also other physical contributions 
beyond QED, to which this imaging method may be applied.

\section{Phase-contrast Fourier imaging}\label{sec2}
We now consider an experimental setup, where we create a
high-intensity spot by focusing a laser pulse in vacuum and
probe its refractive index shift by a second laser pulse.
We call the first laser pulse the target laser pulse,
while from hereon the second laser pulse will be denoted as the probe laser
pulse. We need to detect the extremely small shift of the phase velocity by the
target-probe interaction. For this we also need an intense probe laser
in order to enhance the visibility.
However, if we utilize conventional interferometer techniques, providing a
homogeneous phase contrast over the probe laser profile, such small refractive
index changes are hard to detect. This is because the resulting intensity
modulation always appears on top of a huge pedestal intensity, with an 
extremely small contrast between the modulation and the pedestal.
Any photo-detection device will not be sensitive to the small number of
photons spatially distributed over the pedestal intensity
beyond 1~J ($\sim 10^{18}$ visible photons), due to the limited 
dynamic range of the photon intensity measurable by a camera pixel
without causing saturation of the intensity measurement.
On the other hand, broadening the dynamic range by lowering the gain of
the electric amplification of photo-electrons degrades the sensitivity
to the small number of the spatially distributed photons or the sensitivity to
the small phase shift.
Therefore, we need to invent a method that can spatially separate the weakly
modulated characteristic intensity pattern from the strong pedestal.

\begin{figure}
\includegraphics[width=1.0\linewidth]{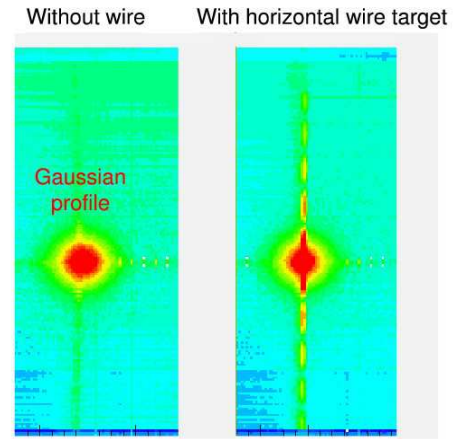}
\caption{
Far-field diffraction patterns from a thin-wire target, 
when a Gaussian laser pulse is shot onto the wire.
The left figure describes the situation without wire.
The right figure shows the case when the thin-wire target is horizontally arranged.
}
\label{Fig2}
\end{figure}

In order to overcome this difficulty,
we suggest utilizing the inhomogeneous phase-contrast Fourier imaging
in the focal plane by focusing the probe laser.
The physically embedded phase contrast on the transverse profile
of the probe laser amplitude is Fourier transformed onto the focal plane
due to the effect of the added phase by the lens.
Actually, a parabolic mirror is necessary to avoid dispersion and
damage by high-intensity irradiation. This will be considered later.
The intensity pattern in the focal plane exhibits the preferable
feature, that the characteristic phase boundary causes outer regions of the 
intensity profile far from the focal point to expand, whereas a Gaussian laser
beam with a homogeneous phase converges into a small focal spot at its waist. 
It is instructive to illustrate the characteristic nature
of the diffraction pattern from a wire-like target as shown in
Fig.~\ref{Fig2}. Here the far-field pattern, known as
Fraunhofer diffraction, is shown in the case when a Gaussian laser beam
irradiates a thin-wire target.
This can be understood as the Fourier
transform of the wire shape, approximated as a rectangle of
$2\mu \times 2\nu$. It is well known that a lens
produces a far-field diffraction pattern, corresponding to
the exact Fourier-transformed image of the object
in the front focal plane (e.g. see \cite{SIEGMAN,Yariv}).
In order to understand the diffraction image, we may qualitatively refer to
Babinet's principle, which states that the diffraction pattern from an
opaque wire plus that of a slit of the same size and shape form
an amplitude distribution identical to that of the incident wave.
Therefore, the characteristic diffraction patterns from the wire and 
the slit are similar, but deviate from each other such that
they interfere to reconstruct the incident wave.
The intensity pattern after Fourier transform of such a rectangular slit 
is expressed as
%
\begin{eqnarray}
   \left(\frac{\sin(\mu\omega_x)}{\mu\omega_x}\right)^2
   \left(\frac{\sin(\nu\omega_y)}{\nu\omega_y}\right)^2,
\end{eqnarray}
%
where $\omega_x=\frac{2\pi}{\lambda f}x$ and $\omega_y=\frac{2\pi}{\lambda f}y$
are the spatial frequencies for the given position $(x,y)$ in the focal plane
of the lens/mirror
with the focal length $f$ at the wavelength $\lambda$, respectively.
In the case of a slit with $\mu\gg\nu$,
the rectangular profile in the focal plane 
emerges as a pattern of dark and bright fringes perpendicular to the slit
(see Fig.~\ref{Fig2}~(right)). 
The narrower the slit size is, the further the fringes move apart.
On the other hand, a Gaussian beam without wire or slit remains unchanged,
because the Fourier transform of a Gaussian beam remains a Gaussian beam 
(see Fig.~\ref{Fig2}~(left)).
This is the key feature that drastically improves the detectability of
small phase shifts by sampling only outer parts of the diffraction pattern.
This may also be interpreted as the counter-concept to the conventional
spatial filter, where outer parts are eliminated to maintain a smooth
phase on the transverse profile of the Gaussian distribution.

Given the intuitive picture above,
a quantitative formulation of our proposed method is presented as follows.
In order to discuss the amount of the phase shift, we need a distinct
geometry of both the target and probe lasers.
Let us first consider the laser profile assuming Gaussian beams.
The solution of the electromagnetic field
propagation along $z$ in vacuum is well-known~\cite{Yariv}.
The electric field component corresponding to the transverse mode 
$l,m$ and {\it e.g.} polarized along $y$ is expressed as 
$\vec{E}(x,y,z,t) = Re\{\vec{e}_y \psi_{l,m}(x,y,z)e^{i\omega t}\}$ with
%
\beqa\label{eq_Hermite}
\psi_{l,m} (x,y,z) = A_0 \frac{w_0}{w(z)}
H_l \left( \frac{\sqrt 2}{w(z)} x \right) 
H_m \left( \frac{\sqrt 2}{w(z)} y \right) \times \nnb\\
\exp 
\left\{
-i[kz-(l+m+1)\eta(z)] - r^2 \left( \frac{1}{{w(z)}^2}+\frac{ik}{2R(z)} \right)
\right\},
\eeqa
%
where the $H_l$ are $l$-th order Hermite polynomials, 
$k=2\pi/\lambda$, $r=\sqrt{x^2+y^2}$, $w_0$ is the waist,
which cannot be smaller than $\lambda$ due to the diffraction limit, and
other definitions are as follows:
%
\beqa\label{eq_wz}
{w(z)}^2 = {w_0}^2
\left(
1+\frac{z^2}{{z_R}^2}
\right),
\eeqa
\beqa\label{eq_Rz}
R = z
\left(
1+\frac{{z_R}^2}{z^2}
\right),
\eeqa
\beqa\label{eq_etaz}
\eta(z) = \tan^{-1}
\left(
\frac{z}{z_R}
\right),
\eeqa
\beqa\label{eq_z0}
z_R \equiv \frac{\pi{w_0}^2}{\lambda}.
\eeqa
%
In order to determine the normalization factor $A_0$,
we use the orthonormal condition of the $n^{th}$ Hermite polynomial as follows
\beq
\int^{+\infty}_{-\infty}{H_n(\xi)}^2 e^{-\xi^2} d\xi = 2^n n! \sqrt{\pi}.
\eeq
With the replacement $\xi = \frac{\sqrt 2}{w_0} x$ we obtain
\beqa
\int^{+\infty}_{-\infty}{H_l^2\left(\frac{\sqrt 2}{w_0} x\right)}
e^{-\frac{2 x^2}{w^2_0}}dx 
 \int^{+\infty}_{-\infty}{H_m^2\left(\frac{\sqrt 2}{w_0} y\right)}
e^{-\frac{2 y^2}{w^2_0}}dy \nnb\\
= \frac{1}{2}{w_0}^2\pi  2^{l+m} l! m! \hspace{1cm}.
\eeqa
At $z=0$ we can then relate the amplitude $A_0$ with the beam power 
$P_0$ by 
\beq\label{eq_Intensity}
P_0 = A^2_0\int^{+\infty}_{-\infty}\int^{+\infty}_{-\infty}
\psi_{l,m}(x,y)\psi^*_{l,m}(x,y) dx dy
\eeq
yielding
\beq
A^2_0 \equiv \frac{P_0}{2\pi\sigma^2 2^{l+m} l! m!},
\eeq
where $w_0$ is replaced by $2\sigma$ and
$A^2_0$ is the on-axis intensity at the waist.

\begin{figure}
   \includegraphics[width=1.0\linewidth]{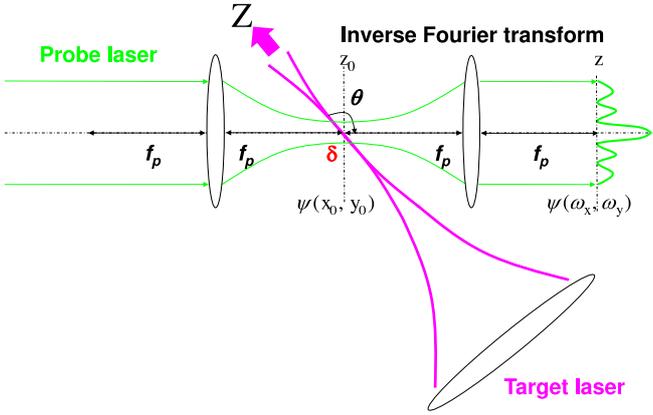}
     \caption{Conceptual experimental setup for the suggested phase-contrast
      Fourier imaging. At the crossing point between the probe and target
      laser beams, the target pulse causes a shift in the index of refraction,
      and a corresponding phase shift $\delta$
      embedded into the probe pulse as explained in Fig.~\ref{Fig3}.
      }
\label{Fig04}
\end{figure}

Figure \ref{Fig04} illustrates the conceptual experimental
setup for the phase-contrast Fourier imaging.
In what follows, subscripts $p$ and $t$ always denote the {\it probe} 
and {\it target} quantities, respectively. 
Both the target and probe laser beams are focused with different
waist sizes ${w_t}_0$ and ${w_p}_0$
with the incident beam diameters $d_t$ and $d_p$, respectively.
Both laser beams cross each other at their waists, where
wave fronts are close to flat with $R=\infty$
in Eq.~(\ref{eq_Rz}).
We assume that the target waist ${w_0}_t$ is smaller than
the probe waist ${w_0}_p$, which embeds the phase contrast $\delta$ at $z_0$
within the amplitude on the transverse profile of the probe laser.
The probe laser then propagates to a lens of 
focal length $f_p$ and the inverse Fourier imaging is performed 
in the back focal plane of that lens.

\begin{figure}
\includegraphics[width=1.0\linewidth]{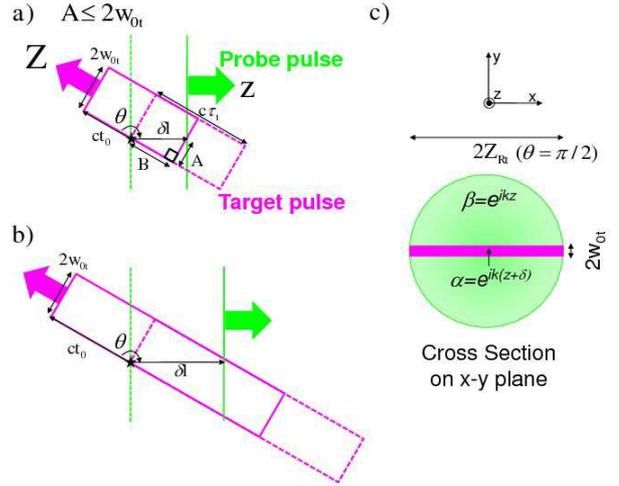}
\caption{Geometry of the embedded phase contrast in the probe pulse.}
\label{Fig3}
\end{figure}

We then define the geometry of the laser intersection. 
Figure~\ref{Fig3} illustrates geometrical relations
where the tightly focused target pulse with time duration $\tau_t$,
beam waist diameter $2w_{0t}$, and Rayleigh length $z_{Rt}$
propagates along the $Z$-axis, and the probe pulse with the larger 
beam waist diameter $2w_{0p}$ and longer time duration $\tau_p$ propagates 
along the $z$-axis tilted by $\theta$ with respect to the $Z$-axis. 
In Fig.\ref{Fig3} a) and b) the pulses 
are assumed to have a rectangular intensity profile along 
the propagation direction. The dashed rectangular pulses occupy the
positions at time $t=0$ and the solid ones are those at $t=t_0$.
The probe and target pulses overlap each other 
at the position marked by $\ast$. At this moment in time, $t$ is taken as zero.
We need to express $\delta l$ to estimate the pass length where
an additional phase is embedded. The path length $\delta l$
is defined as the distance where the front of the probe pulse 
meets the edges of the target laser at $t=t_0$, 
beyond which the target laser is no longer present.
In Fig.\ref{Fig3} a) because $A$, $B$, and $\delta l$ form a right triangle, 
we hence obtain the following relation
\beq\label{eq_triangle}
(\delta l)^2 = A^2 + B^2,
\eeq
where $\delta l = ct_0$, $A=c(\tau_t-t_0)$ and $B=\tan(\pi-\theta)c(\tau_t-t_0)$
with the velocity of light $c$, resulting in
\beq\label{eq_triangle}
t_0 = \frac{\tau_t}{1+\cos(\pi-\theta)}.
\eeq
Depending on the relation between the target beam waist $2w_{0t}$ and
the pulse length $c\tau_t$, namely, whether a) $A \le 2w_{0t}$ or b)
where the probe wavefront meets the side of the target laser before
reaching the tail as shown in Fig.\ref{Fig3},
the optical path length $\delta l$ is expressed as
\begin{description}
\item[a)]
\beqa\label{eq_delta_l_a}
\delta l = c t_0 =
\frac{c\tau_t}{1-\cos\theta}
\mbox{  for } c\tau_t < 2 w_{0t} \frac{1-\cos\theta}{\sin\theta},
\eeqa
\item[b)]
\beqa\label{eq_delta_l_b}
\delta l =
\frac{2 w_{0t}}{\sin\theta}
\mbox{  for } c\tau_t \ge 2 w_{0t} \frac{1-\cos\theta}{\sin\theta},
\eeqa
\end{description}
where the equations should not be applied to the cases 
$\theta=0$ or $\vartheta=\pi$.
In the case a) the path lengths with phase shift are not constant
below or above the star point along the wavefront of the probe laser. 
In the case b) the path lengths are constant over the part of
the probe wavefront which penetrates both sides of the target laser.
The residual part along the wavefront, however, meets the head or
tail of the target laser pulse and causes deviations from the constant
path length.
An exactly equal path length over the probe wavefront during the propagation
time of the target laser $\sim 2 z_{Rt}/c$ is realized only in the case 
of $\theta = \pi/2$. 
In this case, after the penetration of the probe laser pulse, 
the profile of the probe laser in the $x-y$ plane contains a trajectory 
with a constant phase shift $\delta$ along the projection of the path 
of the target laser on the probe wavefront, as shown in Fig.~\ref{Fig3}~c).

We express the phase shift $\delta$ in the vicinity of the waist
$|z|=c\tau_t/2 \le z_{Rt}$, where we assume that the wavefront is flat
as indicated by Eq.~(\ref{eq_wz}) and Eq.~(\ref{eq_Rz})
%
\beqa\label{eq_delta}
\delta = \frac{2\pi}{\lambda_p} \delta n \delta l \varphi_t(x_p,y_p),
\eeqa
where 
$\delta n$ is
the refractive index shift, $\delta l$ is the path length with an
effectively constant phase shift over the crossing time and
$\varphi_t(x_p,y_p)$ is a weighting function to reflect the path length 
difference depending on the incident position with respect to the target 
profile expressed as a function of the position $(x_p, y_p)$ in the transverse
plane of the probe laser.
If we limit the origin of the laser-induced refractive index change to QED,
based on Eq.~(\ref{eq_phsv}), (\ref{eq_zk}), and (\ref{eq_zk_epsilon}),
we parametrize the refractive index shift as
\beqa\label{eq_delta_para}
\delta n_{qed} = \zeta N_0 (1-\cos\theta) \frac{E_t}{\pi {w^2_0}_t c\tau_t},
\eeqa
where
$\zeta$ is 4 or 7 for the polarization combinations $\parallel$ or $\perp$,
respectively.
$N_0$ is the coefficient to convert from energy density to the
refractive index shift defined as
$N_0 \equiv \frac{2}{45}\frac{\alpha^2\hbar^3}{m^4_e c^5}
= 1.67\times 10^{-12}  [\mu\mbox{m}^3/\mbox{J}]$.
The incident angle $\theta$ varies from 0 to $\pi$ which
is measured from the propagation direction of the target pulse to
that of the probe pulse as depicted in Fig.~\ref{Fig3}.
$E_t$ is the energy of the target pulse given in [J], and
$\pi {w^2_0}_t c\tau_t$ is the volume in [$\mu$m${}^3$]
for the given target profile
with the waist ${w_0}_{t}$ from Eq.~(\ref{eq_wz}).

By respecting the constant path length over $\sim 2z_{Rt}/c$ for simplicity,
we consider only the case of (\ref{eq_delta_l_b}) with $\theta = \pi/2$.
By substituting Eq.~(\ref{eq_delta_para}) and (\ref{eq_delta_l_b}) into
Eq.~(\ref{eq_delta}), we obtain the simplest expression for $\delta_{qed}$
\beqa\label{eq_delta_exp}
\delta_{qed} \sim
4\zeta N_0 \frac{E_t}{\lambda_p {w_0}_{t} c\tau_t},
\eeqa
where $c\tau_t \ge 2{w_0}_{t}$ must be satisfied from Eq.(\ref{eq_delta_l_b}) 
and we take the approximation $\varphi_t(x_p,y_p) \sim 1$ 
to simplify the following argument
(if necessary, we may restore the target profile $\varphi_t(x_p,y_p)$
based on the precise profile of the target laser reflecting actual experimental
setups). In this limit we approximate the target profile as
a rectangular of the size $2\mu \times 2\nu$, inside which
the phase shift is assigned to be constant. 
The effective slit sizes are defined by
the transverse sizes of the focused laser beams through the relation
\beq\label{eq_munu}
\mu \sim {z_R}_t \mbox{ and } \nu \sim {w_0}_t.
\eeq
We then explicitly define the window functions $rec$ and $\overline{rec}$ as
%
\begin{eqnarray}\label{eq_slit}
rec(\mu, \nu) = \left\{
\begin{array}{ll}
1 & \quad \mbox{for $|x|\le\mu \mbox{ and } |y|\le\nu$} \\
0 & \quad \mbox{for $|x|>\mu \quad\mbox{or } |y|>\nu$}
\end{array}
\right\},
\nnb\\
\overline{rec}(\mu, \nu) = \left\{
\begin{array}{ll}
0 & \quad \mbox{for $|x|\le\mu \mbox{ and } |y|\le\nu$} \\
1 & \quad \mbox{for $|x|>\mu \quad\mbox{or } |y|>\nu$}
\end{array}
\right\}.
\end{eqnarray}
%
This window provides a unit region of a constant phase, which may be
applied even to arbitrary phase maps composed of a collection of
the unit window cells. 

We now discuss how the probe laser including the phase $\delta$ embedded
at the focal plane propagates into the image plane via the lens system
and evaluate the expected intensity distribution in the image plane.
Since our discussion is based on local phases with rectangular shape
and both Fourier and inverse Fourier transforms of a rectangular function
give identical sinc functions, we represent the lens effect as 
Fourier transform. 
For each propagation from the object plane $(x_0, y_0)$ at $z_0$ to the
image plane $(x, y)$ at $z$ (see Fig.\ref{Fig04}), 
we always take the Fresnel diffraction.
Based on (\ref{eq_Hermite}) and (\ref{eq_Intensity}),
the probe field profile in the plane where the phase $\delta$ is
embedded can be defined as 
\beqa\label{eq_P}
T(x_0, y_0) = A_p H_l \left( \frac{x_0}{\sqrt 2 \sigma} \right)
                  H_m \left( \frac{y_0}{\sqrt 2 \sigma} \right) 
                  e^{-\frac{x_0^2+y_0^2}{(2\sigma)^2}}
\eeqa 
where $A_p \equiv \sqrt{\frac{I_0}{2\pi\sigma^2 2^{l+m} l! m!}}$ is
the on-axis waist amplitude of the probe laser.
The linearly synthesized amplitude at $z_0$ is then expressed as
%
\beqa\label{eq_planewave}
\Psi(x_0, y_0, z_0) = \alpha(z_0) rec(\mu,\nu) T(x_0,y_0) + \nnb\\
                 \beta(z_0) \overline{rec}(\mu,\nu) T(x_0,y_0),
\eeqa
%
where $\alpha(z)$ and $\beta(z)$ are propagation factors of the 
probe waves at the point $z$ after probe-target crossing.
The functions $\alpha$ containing the phase shift
$\delta$ caused by the local refractive index shift and $\beta$ are defined as
%
\begin{eqnarray}\label{EqPlane}
\alpha(z) &=& e^{i(kz+\delta)}, \nonumber \\
\beta(z)  &=& e^{ikz}.
\end{eqnarray}
%
The Fourier transform $F$ of the synthesized amplitude $\Psi$ in
the image plane $(x,y)$ at $z$ after the lens~\cite{Goodman} is expressed as
%
\begin{eqnarray}\label{EqF}
F\{\Psi(x_0, y_0)\} =
\qquad \qquad \qquad \qquad \qquad \qquad \qquad \qquad \qquad \nnb\\
\alpha(z_0) F\{rec(\mu,\nu) T(x_0,y_0)\} +
\beta(z_0) F\{\overline{rec}(\mu,\nu) T(x_0,y_0)\} \nnb\\
= (\alpha(z_0)-\beta(z_0)) \int^{\mu}_{-\mu}\!\!\int^{\nu}_{-\nu}\!\!dx_0dy_0
T(x_0,y_0) e^{-i(\omega_x x_0 + \omega_y y_0)} +
\nnb\\
\beta(z_0) \int^{\infty}_{-\infty}\!\!\int^{\infty}_{-\infty}\!\!dx_0dy_0
T(x_0,y_0) e^{-i(\omega_x x_0 + \omega_y y_0)},\hspace{0.4cm}
\end{eqnarray}
%
where we define $(\omega_x,\omega_y)\equiv(\frac{2\pi}{f_p\lambda_p}x,
\frac{2\pi}{f_p\lambda_p}y)$ at $z$. We introduce the coefficient $C_{s}$
for the first term in Eq.~(\ref{EqF}),
containing the information on how much the phase shift,
representing the {\it signal}, is localized, 
resulting in the photon-photon interaction.
We decompose $C_{s}$ into its real and imaginary parts, because
Hermite polynomials contain even and odd functions and the non-zero
values of these integrals appear even in the imaginary part.
We denote them as
%
\begin{eqnarray}
\mbox{Re} C_{s}(\omega_x, \omega_y) \equiv C_{sR} =
 \qquad \qquad \qquad \qquad \qquad \qquad \qquad \nnb\\
 \int^{\mu}_{-\mu}\!\!dx_0 \int^{\nu}_{-\nu}\!\!dy_0
 \{\cos(\omega_x x_0) \cos(\omega_y y_0) - \nnb\\
   \sin(\omega_x x_0) \sin(\omega_y y_0)\}T(x_0,y_0)\nnb
\end{eqnarray}
and
\begin{eqnarray}\label{EqCsig}
\mbox{Im} C_{s}(\omega_x, \omega_y) \equiv C_{sI} =
 \qquad \qquad \qquad \qquad \qquad \qquad \qquad \nnb\\
-\int^{\mu}_{-\mu}\!\!dx_0 \int^{\nu}_{-\nu}\!\!dy_0
 \{\cos(\omega_x x_0) \sin(\omega_y y_0) + \nnb\\
   \sin(\omega_x x_0) \cos(\omega_y y_0)\} T(x_0,y_0). \qquad
\end{eqnarray}
%
We also define the coefficient $C_{b}$ for the second term of Eq.~(\ref{EqF}),
which corresponds to the {\it background} pedestal as
%
\begin{eqnarray}\label{EqCbkg}
C_{b}(\omega_x, \omega_y) \equiv
\quad \qquad \qquad \qquad \quad
\quad \qquad \qquad \qquad \nnb\\
(-i)^{l+m} (2\sigma)^2\pi 
H_l(\sqrt 2 \sigma \omega_x) H_m(\sqrt 2 \sigma \omega_y)
e^{-\frac{\omega_x^2+\omega_y^2}{(2\sigma)^2}},
\end{eqnarray}
where the fact is used that the $n_{th}$-order Hermite function is 
the eigen-function of the Fourier transform, namely, 
$(2\pi)^{-1/2} F\{H_n(\xi) e^{-\xi^2/2}\} =
(-i)^n H_n(\omega) e^{-\omega^2/2}$.
We note that the coefficient $(2\pi)^{-1/2}$ arises due to 
our definition of the Fourier transform with the prefactor of unity 
applied to the lens system. 
The Fourier transform of the amplitude is then expressed as
%
\begin{eqnarray}\label{EqFfinal}
F\{\Psi\} =
(\alpha-\beta) (C_{sR} + i C_{sI})
+ \beta (C_{bR} + i C_{bI}).
\end{eqnarray}
%
When no confusions are expected,
we will  omit $(x_0, y_0, z_0)$ and $(\omega_x, \omega_y, z)$ for
all relations below (\ref{EqFfinal}).
By substituting Eq.~(\ref{EqPlane}), (\ref{EqCsig}) and (\ref{EqCbkg})
into Eq.~(\ref{EqFfinal}), the intensity pattern at the image plane
is expressed as
%
\beqa\label{eq_focalint}
|F\{\Psi\}|^2 = \psi_{l,m} \psi^*_{l,m} =
\left( \frac{A_p}{f_p\lambda_p} \right)^2 \times \hspace{2.5cm}\nnb\\
\{ 
 2(1-\cos\delta) \left( C_{s}C^*_{s} - ( C_{sR} C_{bR} + C_{sI} C_{bI} ) \right) 
\nnb\\
 -2\sin\delta ( C_{sI} C_{bR} - C_{sR} C_{bI} ) + C_{b}C^*_{b}
\}, \quad
\eeqa
where $(f_p\lambda_p)^{-2}$ arises from the Fresnel diffraction.
According to (\ref{EqCbkg}), when $l+m$ is even or odd, 
$C_{bI}$ or $C_{bR}$ becomes zero, respectively.

Equation (\ref{eq_focalint}) indicates that this method works
as an interferometer via the cross terms with coefficients
$1-\cos\delta$ and $\sin\delta$. 
The second term with $\sin\delta$ vanishes for any combinations of $l$ and $m$,
as long as the symmetric rectangular ranges around $(x_0, y_0) = 0$ are 
assumed in the definitions of Eq.(\ref{EqCsig}).
This interferometer differs from a conventional one in
that the modulating part due to the phase shift $\delta$ can be spatially
separated from the confined strong part $C_{b}C^*_{b}$, due to the
characteristic pattern of $C_{s}C^*_{s}$ which creates images in the
region of higher spatial frequencies.
In Eq.(\ref{eq_focalint})
the third term $C_{b}C^*_{b}$ corresponds to the intense pedestal pattern
insensitive to the phase $\delta$ which keeps the same shape as that at
$z_0$ with a different transverse scale based on Eq.(\ref{EqCbkg}),
because Hermite functions are the eigen-functions of the Fourier transform.
The second term shows proportionality to $\delta$ for $\delta \ll 1$; 
however, the intensity pattern is constrained by $C_{b}$. 
This implies the phase information is attainable only in the vicinity 
of the background pattern $C_{b}C^*_{b}$, though the signal is strong.
Therefore, the signal-to-pedestal ratio is not expected to be large. 
On the other hand, the first term indicates proportionality to $\delta^2$
for $\delta \ll 1$ 
which implies a very weak signal; however, the $C_{s}C^*_{s}$ term is not 
affected by $C_{b}$ and it produces a pattern characterized by spatial
frequencies.
Hence the signal-to-pedestal ratio 
is expected to be drastically improved circumventing the most intense spot.
Therefore, depending on the value of $\delta$ and the allowed dynamic range 
of the photo-detection device used, we have choices on which term 
we focus.

\begin{figure}
   \includegraphics[width=1.0\linewidth]{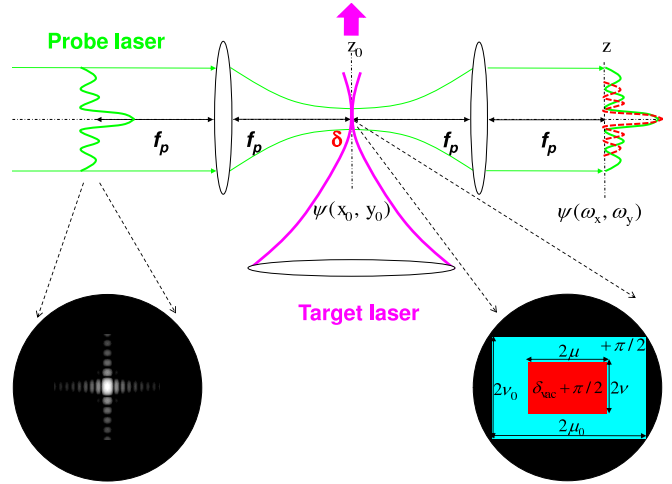}
     \caption{Conceptual experimental setup for the suggested phase-contrast
      Fourier imaging. At the crossing point between the probe and target
      lasers, the target laser causes a shift in the index of refraction,
      which amounts to the refractive phase shift $\delta_{vac}$
      embedded into the probe laser as explained in Fig.~\ref{Fig3}.
      The offset phase is embedded by a holographic phase plate in advance 
      which provides the offset phase $+\pi/2$ inside the rectangular region
      $rec(\mu_0, \nu_0)$ at $z_0$ via Fourier transform by the first lens.
      }
\label{Fig4}
\end{figure}

We consider a Gaussian beam with $l=m=0$.
Therefore, we do not expect the terms proportional to phase $\delta$ due to the
imaginary parts originating from Hermite polynomials with odd orders.
Nevertheless, if we need to stick to the proportionality to $\delta$,
we may add a local offset phase $\pi/2$
along the path of the target laser on the focal plane.
We define this plane as the object plane for the inverse Fourier imaging
as illustrated in Fig.\ref{Fig4}.
Such a setup can recover the sensitivity to 
the sign of the phase shift as well as the absolute value, 
because $1-\cos(\delta+\pi/2)$ in Eq.~(\ref{eq_focalint}) 
becomes $\sim 1+\delta$. It is important to be able to discuss whether the
phase shift is increased or decreased, since it directly
reflects the dynamics of the local interaction. From a technical point of view,
more importantly, this has a definite advantage of enhancing the signal
due to the proportionality to $\delta$ compared to the $\delta^2$ sensitivity
in $1-\cos\delta$ in case of an extremely small $\delta$. However, in turn,
one must accept the situation that the local offset phase contrast 
produces an intrinsic diffraction pattern as a new kind of pedestal, which
now has an equal diffraction pattern compared to the one caused by 
the photon-photon interaction. Thanks to the proportionality to $\delta$, 
we can reduce the intensity of the probe laser pulse.  
On the other hand, the new pedestal pattern would occupy the dynamic 
range of the camera device.
In order to reduce the amount of the pedestal intensity,
we may add more intelligent characteristic offset patterns 
by mixing $\delta \equiv \delta_{vac} \pm \pi/2$ 
with the laser-induced vacuum phase shift $\delta_{vac}$
and the offset phase $\pm \pi/2$
so that the offset diffraction patterns 
can destructively interfere in some points in the image plane 
thus still keeping a high signal-to-pedestal ratio.
Therefore, the implementation of such local offset phases on the probe laser
in advance is a key design issue, depending on the
dynamic range of the camera device. 

As illustrated in the zoom of the object plane 
(focal plane common to both lenses) in Fig.\ref{Fig4},
we consider a rectangular phase offset so that it
contains the region with phase $\delta_{vac} \pm \pi/2$ in its center, 
that is, we define the offset region as $rec(\mu_0, \nu_0)$ 
with $\mu_0 = N\mu$ and $\nu_0 = N\nu$ and $N \ge 1$ in Eq.(\ref{eq_slit}) 
by giving the offset phase of $\pm \pi/2$.
For this region we introduce the coefficient 
$C_0(\omega_x, \omega_y)$ by replacing $(\mu, \nu)$ with $(\mu_0, \nu_0)$
in Eq.(\ref{EqCsig}) as well. The intensity profile in the focal plane
is then re-expressed as
%
\begin{eqnarray}\label{eq_focalint_offset}
|F \{ \Psi \}|^2 = \psi_{l,m}\psi^*_{l,m} \sim
\left( \frac{{A_0}_p}{f_p\lambda_p} \right)^2 \times
\quad \qquad \qquad \nnb\\
\{ 2(C_0-C_{b})(C_0 \pm \delta_{vac} C_{s}) + C_{b}C^*_{b} \}, \quad
\end{eqnarray}
%
where $\pm$ are cases when the offset phase $\pm\pi/2$ are added
in $rec(\mu_0, \nu_0)$ and $\overline{rec}(\mu_0, \nu_0)$, respectively.
We note that this relation is applicable to both real and imaginary 
coefficients as long as either all real or all imaginary coefficients 
are simultaneously zero.
Actually we can confirm that Eq.(\ref{eq_focalint}) becomes identical 
with Eq.(\ref{eq_focalint_offset}) under this condition,
when $N=1$ and $\delta_{vac} = 0$, namely, 
$C_s = C_0$ and $\delta = \pm \pi/2$ are substituted into 
Eq.(\ref{eq_focalint}).

\begin{figure}
\includegraphics[width=1.0\linewidth]{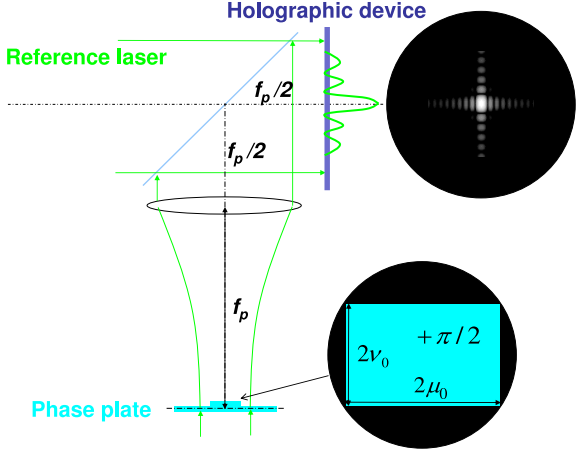}
\caption{An example how to implement the offset phase by a holographic device.}
\label{hologram}
\end{figure}

The offset phase may be embedded by a sinc distribution 
which is, for example, produced via a step-like phase 
plate in advance 
via Fraunhofer diffraction by locating the plate at a far distance 
from the first lens in Fig.\ref{Fig4}.
This may provide the offset phase $+\pi/2$ within
the rectangular region $rec(\mu_0, \nu_0)$ at $z_0$ via 
Fourier transform by the first lens in Fig.\ref{Fig4}.
If such a long distance is not available, we may use a holographic device
as illustrated in Fig.\ref{hologram} where the phase plate is placed
at the focal plane and a laser produces a proper Fourier image which
is stored in the holographic device by mixing with a reference laser. 
If we replace the reference laser by the probe laser in Fig.\ref{Fig4},
we can produce the Fourier image in front of the first lens
in Fig.\ref{Fig4}. In a practical case as shown in
Fig.\ref{Fig7}, the holographic device may be located before the
focal point when it records the local phases in advance,
in order to supply an offset distance before the proper sinc distribution is 
reconstructed by probe laser pulses at the exact point where we need it.

\section{Analysis in the Fourier image}\label{sec3}
\begin{table*}[t]
\begin{tabular}{lr}
\hline\hline
{\bf Target laser parameters} & {\bf Probe laser parameters} \\ \hline\hline
$\tau_t = 15$~fs & 
$\tau_p = 2{z_R}_t/c = 24$~fs\\ \hline
$E_t=250\mbox{J}$ 
&
$E_p=25\mbox{J}$ 
\\ \hline
$\lambda_t = 800 \pm 40$~nm & $\lambda_p = 800 \pm 40$~nm \\ \hline
$d_t = 39.8$~cm & $d_p = 7.0$~cm  \\ \hline
$f_t = 75$~cm & $f_p = 25$~cm  \\ \hline
${w_0}_t \sim \frac{2f_t \lambda_t}{\pi d_t}= 0.96\mu$m & 
${w_0}_p \sim \frac{2f_p \lambda_p}{\pi d_p}= 1.8\mu$m\\ \hline
${z_R}_t=\pi{w^2_0}_t/\lambda_t = 3.6\mu$m &
${z_R}_p=\pi{w^2_0}_p/\lambda_p = 18.5\mu$m \\ \hline\hline
{\bf Embedded physical phase by assuming only QED effect}\\ \hline\hline
$\delta_{qed} = 1.28 \times 10^{-10}$ from Eq.(\ref{eq_delta_exp})
with $\zeta =4$ and $\theta = \pi/2$ in Eq.(\ref{eq_delta_para}) \\ 
\hline\hline 
{\bf Shape of physical and offset phases}\\ \hline\hline
$\mu = {z_R}_t$ and $\nu = {w_0}_t$\\ \hline
$\mu_0 = 5 \times \mu$ and $\nu_0 = 5 \times \nu$ with offset phase $+\pi/2$\\ \hline
\end{tabular}
\caption{
Laser parameters used to produce Fig.~\ref{Fig5} 
based on the conceptual experimental setup shown
Figs.~\ref{Fig3} and \ref{Fig4}.
The subscripts $t$ and $p$ refer to the target and probe lasers, respectively.
This choice of parameters is explained in the text in
sections~\ref{sec2} and \ref{sec3}.
}
\label{Tab1}
\end{table*}

\begin{figure*}
\includegraphics[width=1.00\linewidth]{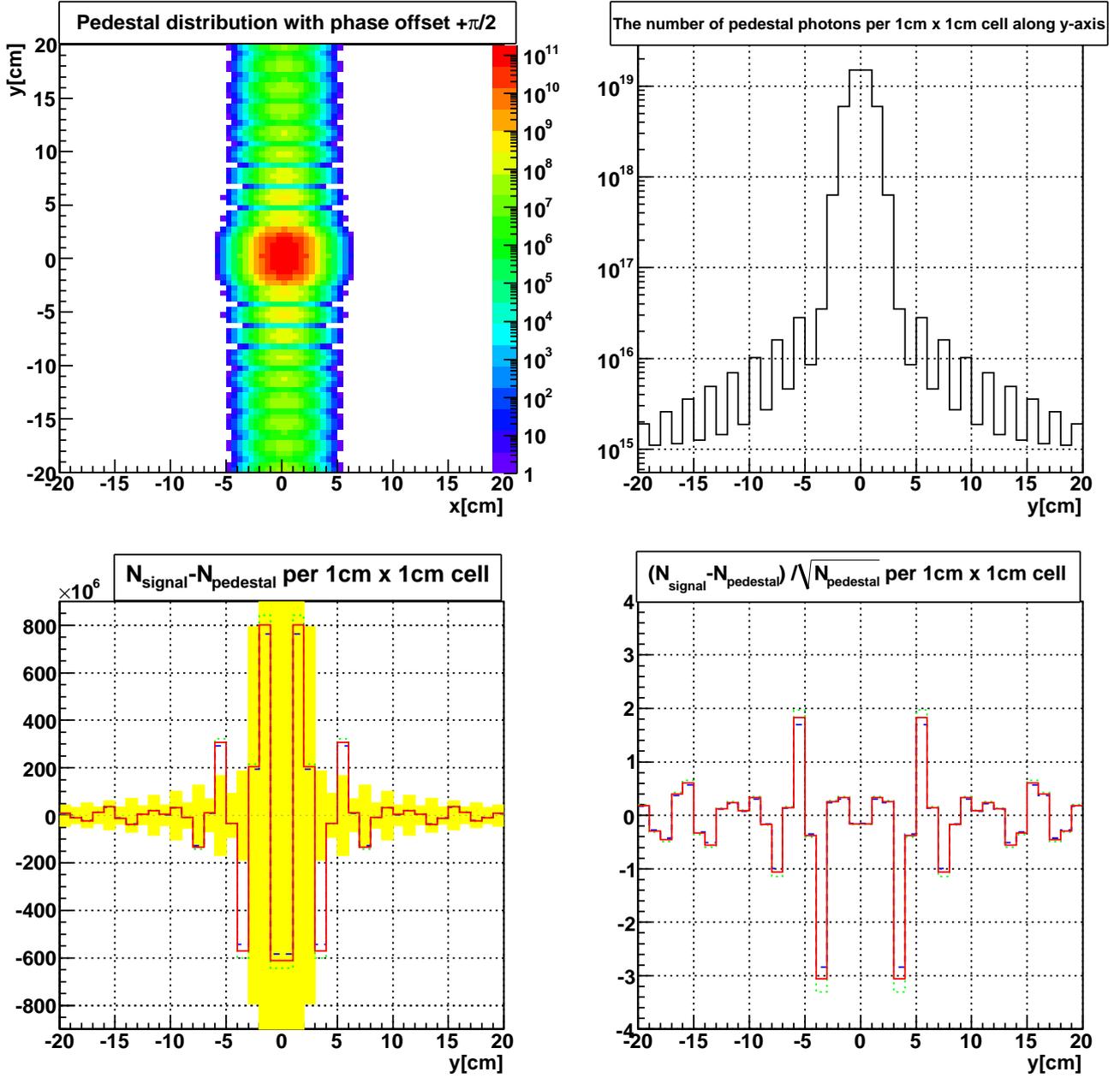}
\caption{
Simulated distributions of $\psi_{0,0}\psi^*_{0,0}$ (TEM${}_{00}$) 
in image plane $z$ based on Eq.(\ref{eq_focalint_offset}) with parameters
given in Tab.~\ref{Tab1}.
{\it top-left}: 
Patterns of Eq.(\ref{eq_focalint_offset}) with the offset phase $+\pi/2$ only.
{\it top-right}:
The number of pedestal photons, $N_{ped}$ integrated over a 1cm x 1cm 
cell along the y-axis at $x=0$ of the {\it top-left} distribution.
{\it bottom-left}: 
$N_{sig} - N_{ped}$ per 1cm x 1cm cell along the y-axis at $x=0$,  
where $N_{sig}$ is the integrated number of photons per 1cm x 1cm 
cell with $\delta = \delta_{vac} +\pi/2$.
The solid-red, dashed-blue, and dotted-green histograms 
in Fig.\ref{Fig5} show the case when 
probe wavelengths of 800nm, 840nm, and 760nm are assumed, respectively,
and the yellow band shows statistical fluctuations, $\sqrt{N_{ped}}$,
due to the quantum efficiency of the photon detector.
{\it bottom-right}: 
The statistical significance of signal photons
with respect to the statistical fluctuations
of the background photons; $(N_{sig}-N_{ped})/\sqrt{N_{ped}}$
per 1cm x 1cm cell along the y-axis at $x=0$.
The colors have the same meaning as those in the {\it bottom-left} chart.
}
\label{Fig5}
\end{figure*}

We performed numerical calculations with the rectangular offset phase
$+\pi/2$ based on the setup illustrated
in Fig.\ref{Fig4} with Eq.(\ref{eq_focalint_offset}) for 
$l=m=0$ (TEM${}_{00}$). 
The parameters used for Fig.~\ref{Fig5} are summarized
in Tab.~\ref{Tab1}, where the parameters of the target and probe lasers,
the embedded offset and the physical phase shifts due to the nonlinear 
QED effect used to obtain the Fourier transformed intensity distributions
are specified.
Figure~\ref{Fig5}~{\it top-left} illustrates the intensity pattern due to
Eq.(\ref{eq_focalint_offset}) as a function $(x, y)$ in the image plane
when $\delta_{vac} = 0$ and the offset phase $+\pi/2$ is embedded only.
The figure is plotted with an arbitrary unit for the contour height 
in logarithmic scale, by sampling values with $5$~mm steps along 
the $x$ and $y$-axes. 
Figure~\ref{Fig5}~{\it top-right} shows the expected number of pedestal
photons, $N_{ped}$ integrated over a 1cm x 1cm cell along the y-axis at $x=0$.
Figure~\ref{Fig5}~{\it bottom-left} shows $N_{sig} - N_{ped}$ 
per 1cm x 1cm cell along the y-axis at $x=0$,  
where $N_{sig}$ is the integrated number of photons per 1cm x 1cm 
cell for signal, namely, with $\delta = \delta_{vac} +\pi/2$. 
In the actual experimental setup the subtraction should be performed on
the shot-by-shot basis as illustrated in Fig.\ref{Fig7} where a probe laser
pulse is equally split into the signal path with the target laser pulse and 
the calibration path without it so that we can compare the two cases. 
The solid-red, dashed-blue and dotted-green histograms show the case 
when the probe wavelengths of 800nm, 840nm, and 760nm are assumed, respectively,
and the yellow band shows the statistical fluctuations $\sqrt{N_{ped}}$
due to the quantum efficiency of the photon detection.
Figure~\ref{Fig5}~{\it bottom-right} shows the statistical significance
of the signal photons with respect to the statistical fluctuations
of the pedestal photons; $(N_{sig}-N_{ped})/\sqrt{N_{ped}}$
per 1cm x 1cm cell along the y-axis at $x=0$.
The colors have the same meaning as those in Fig.~\ref{Fig5}~{\it bottom-left}.

Figure~\ref{Fig5}~{\it bottom-right} indicates that 
we can expect several cells in which 
the number of signal photons is either increased or decreased by
more than two standard deviations
from the pedestal fluctuations. If we could count the number of photons
per 1cm x 1cm cell in the side band around the pedestal peak without
detector saturation, we can measure the phase velocity 
shift due to the QED effect even by one probe-target laser crossing. 
This side band structure appears owing to interference between
$C_0(\omega_x, \omega_y)$ and $C_s(\omega_x, \omega_y)$ 
in Eq.(\ref{eq_focalint_offset}).
 
Although the spectral width of the probe laser shows
faint effects as shown in 
Fig.\ref{Fig5}~{\it bottom-left} and {\it bottom-right},
the characteristic pattern along the y-axis is similar. As long as 
the wavelength distribution can be measured at the same time, 
we can reconstruct $\delta_{vac}$ based on the measured wavelength distribution
and the intensity pattern along the y-axis.
 
The most difficult issue is the dynamic range of existing cameras used in
research which typically have 16-bit resolution and at most 28-bit per pixel.
In order to solve the
limited dynamic range, let us suppose that we sample photons per 1cm x 1cm cell
by $\sim 10^6$ pixels. In such a case the number of photons per pixel is 
$\sim 10^9$ with respect to $\sim 10^{15}$ photons at around 5cm from 
the pedestal peak (see Fig.\ref{Fig5} {\it top-right}).
Even if we use 10-bit resolution, the number of
photons per resolution becomes $10^9/2^{10} \sim 10^6$ photons. Compared to the
$N_{sig}-N_{ped}$ of $\sim 10^8$ at around 5cm from the pedestal peak
(see Fig.\ref{Fig5} {\it bottom-left}), 
the sensitivity of $10^6$ photons per resolution is sufficient to observe
the intensity modulations beyond two standard deviations from the pedestal
fluctuations without intensity saturation 
(see Fig.\ref{Fig5} {\it bottom-right}). 
This suggests that in principle it is possible to detect 
the laser-induced QED effect by a single shot only if the conditions listed
in Tab.\ref{Tab1} are realized.
Therefore, by assigning camera devices for
individual 1cm x 1cm cell with $\sim 10^6$ pixel readout,
we can overcome the limited dynamic range of cameras
even with the currently existing technology.


In order to study the laser-induced vacuum birefringence,
we inject a linearly polarized probe pulse 
whose electric field vector is turned by 45 deg with respect to
that of the target pulse so that its electric field along the $x$ and $y$
axes are equal. 
We then put two polarization filters at the image plane
symmetrically with respect to $y=0$ as illustrated in Fig.\ref{Fig7}
to cover the regions $+y$ and $-y$ along the $y$-axis, respectively, 
which select orthogonal polarizations at the image plane. 
The asymmetry between the number of modulated
photons from that of the pedestal pattern between the regions $\pm y$
provides direct information of the birefringence on the pulse-by-pulse basis.


We note that this method bears similarity to that in \cite{NaturePhotonics},
where two intense target laser pulses are treated as a matterless double
slit and the interference between spherical waves from these slits
is discussed as a signature of the photon-photon interaction.
In \cite{NaturePhotonics} the occurrence of diffraction is caused by the
laser-laser interaction itself.
In our method the target laser causes the refractive
phase shift experienced by the probe laser, as indicated in Fig.~\ref{Fig4}.
This phase shift is embedded in a refracted, nearly-plane wave
in the forward direction of the probe laser,
as explicitly formulated in Eq.~(\ref{eq_planewave}) and Eq.~(\ref{EqPlane}).
We then set a lens to the right of the interaction between the target and
probe lasers as shown in Fig.~\ref{Fig4}.
The diffraction or Fourier transform in our method is incurred
by the added phase of the lens
and the spherical wave propagation from the lens to the focal plane.
The advantage of our method is an enhanced sensitivity to a small phase shift
on the pulse-by-pulse basis, as it is demonstrated
due to a more efficient collection of photons by the lens
using the much simpler target geometry.
On the other hand, the disadvantage is the deviation from the ideal phases
included in the path of the probe laser except the laser-induced
vacuum phase.
The ways to correct for this kind of background phase aberrations 
and the other background source for the phase-contrast Fourier imaging
will be discussed in the following subsections.

\subsection{Template analysis for local phase reconstruction}\label{subsecA}
\begin{figure}
   \includegraphics[width=1.0\linewidth]{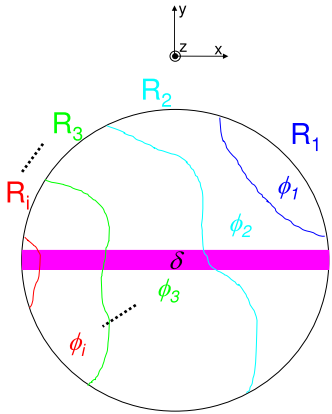}
   \caption{The phase $\delta$ induced by the target laser pulse
            in the presence of local phase fluctuations 
            in the transverse plane of the probe laser pulse 
            at the object plane 
            where the target and the probe pulses cross each other.
            }
\label{Fig6}
\end{figure}

In actual experiments it is unavoidable that the probe pulse includes
local phase fluctuations on a pulse-by-pulse basis 
even in the absence of the laser-induced signal $\delta$
as illustrated in Fig.\ref{Fig6}.
The figure corresponds to the case when the phase contrast $\delta$
in Fig.\ref{Fig4}~c) is embedded in the presence of background phase 
fluctuations $\phi_i \equiv \phi(X)$ as a function of the position 
$X \equiv (x_0, y_0)$ at the object plane
where $i$ denotes a corresponding region
with the constant phase $\phi_i$ in the transverse plane of the probe pulse.
Compared to $\delta$, the $\phi_i$'s are expected to be much larger. 
However, if the values of the local phase set $\{\phi_i\}$ on each probe 
pulse is suppressed below the offset phase $\pi/2$ and the set is 
{\it a priori} measured, we are in principle able to correct for the effect of 
the background fluctuations.
In the next subsection \ref{subsecB} we discuss how to measure
the phase set on a pulse-by-pulse basis in detail.
In this subsection, however, we focus on how to determine 
$\delta$ on a pulse-by-pulse basis,
if the measured phase set is given in advance.

Let us extend the expressions from Eq.~(\ref{eq_slit}) through (\ref{EqF}).
In general, the integration limits defined by Eq.~(\ref{eq_slit}) and
used in the first equation of (\ref{EqF})
can take any shape and size. We replace the rectangular
region $rec$ with the region $R_i \equiv R(X)$,
where a constant phase is mapped within $R_i$.
By denoting the spatial frequency as
$W=(\omega_x,\omega_y) = (2\pi x/(f_p\lambda_p), 2\pi y/(f_p\lambda_p))$,
for the position $(x, y)$ at the image plane with the integral kernel
$f(W, X) \equiv T(x_0, y_0) e^{-i(\omega_x x_0 + \omega_y y_0)}$,
the Fourier transform including the local phase fluctuations $\phi_i$ is
expressed as
%
\begin{eqnarray}\label{eq_psiW}
\Psi(W; \phi) &=& F\{\Psi(X; \phi)\} \nnb\\
&=& \sum^{N_X}_i \{\alpha(\phi_i)-\beta\} \int_{R_i} dX f(X,W) \nnb\\
&=& \sum^{N_X}_i \{\alpha(\phi_i)-\beta\} {\cal F}_i(W) + \beta {\cal F}_{\infty}(W),
\end{eqnarray}
%
where $N_X$ is the number of regions in the transverse plane at $z$,
$\alpha(\phi_i) = e^{i(kz + \phi_i)}$, $\beta = e^{ikz}$,
${\cal F}_i(W) = \int_{R_i} dX f(X,W)$, and
${\cal F}_{\infty}(W) = \int^{\infty}_{-\infty} dX f(X,W)$.
We note that this expression corresponds to the regional cut and paste
on $T(x_0, y_0)$; {\it i.e.},
cutting a region with a phase determined from
$\beta$ at $z$ and paste the same region by adding $\phi_i$ in
$\alpha(\phi_i)$.

Given $\phi_i$ on a pulse-by-pulse basis,
we can numerically calculate the
real and imaginary parts of $\Psi(W;\phi)$.
The estimated background intensity pattern $I_{bg}(\phi)$
in the image plane with the phase fluctuations $\phi$
without the laser-induced phase is given by
\begin{eqnarray}\label{EqIphi}
I_{bg}(W;\phi) = \{\mbox{Re} \Psi(W;\phi) \}^2 + \{\mbox{Im} \Psi(W;\phi) \}^2.
\end{eqnarray}

We now include as well a template of the laser-induced phase by the target
laser pulse $\delta \equiv \delta(X)$. The phase shift $\delta$ can be
evaluated from the geometry of the energy density profile of the target laser
pulse. Based on Eq.(\ref{eq_delta}) we parametrize $\delta$ as
\beq\label{eq_kappaphi}
\delta = \kappa \varphi_t(X),
\eeq
where $\kappa$ is a constant parameter that considers the absolute value
of the phase shift induced by the target laser.
The profile can be {\it a priori} determined by the experimental design 
of the focal spot. 
We can monitor if the center of the spot is in fact stable
and further correct for its deviation from the fixed geometry 
of the target laser. Given $\delta$, we only have to replace the phase
by $\phi_i \rightarrow \phi_i + \delta$ with a constant
parameter $\kappa$ as follows
%
\begin{eqnarray}\label{EqIphidelta}
I_{bg+sig}(W;\phi+\kappa\varphi_t) =
\qquad \qquad \qquad \qquad \qquad \nnb\\
\{\mbox{Re} \Psi(W;\phi+\kappa\varphi_t) \}^2 + \{\mbox{Im} \Psi(W;\phi+\kappa\varphi_t) \}^2,
\end{eqnarray}
%
where $bg+sig$ refers to the fact that the laser-induced phase is 
embedded in the background phase fluctuations.

Given the measured intensity pattern $I_{meas}$ in the image plane
per probe pulse,
we define $\chi^2$ with Eq.~(\ref{EqIphidelta}) as a function of $\kappa$
%
\begin{eqnarray}\label{eq_chi2}
\chi^2(\kappa) \equiv
\qquad \qquad \qquad \qquad
\qquad \qquad \qquad \qquad \qquad \nnb\\
\frac{1}{N_W-1} \sum_j^{N_W}
\frac{|I_{meas}(W_j)-I_{bg+sig}
(W_j;\phi+\kappa\varphi_t)|^2}{I_{meas}(W_j)+I_{bg+sig}(W_j;\phi+\kappa\varphi_t)},
\end{eqnarray}
%
where $N_W$ is the number of sampling points in the image plane and
$j$ runs over all regions in this plane.
The parameter $\kappa$ can be determined by minimizing $\chi^2$
on a pulse-by-pulse basis within the required accuracy.

\subsection{Corrections on pulse-by-pulse phase aberrations}\label{subsecB}
\begin{figure}
   \includegraphics[width=1.0\linewidth]{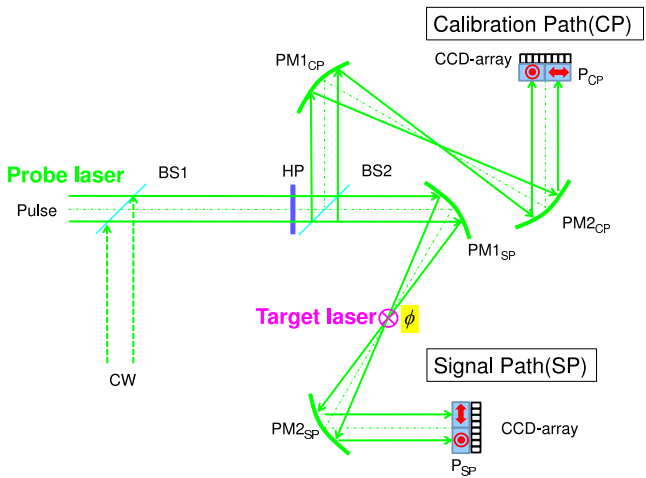}
   \caption{Setup to correct local phase fluctuations.}
\label{Fig7}
\end{figure}

Figure \ref{Fig7} illustrates a schematic view of the entire system
for the phase-contrast Fourier imaging including parts to correct all
phase aberrations in the system.
The target laser pulse moves perpendicular to the drawing plane.
Its focus or waist lies in this plane.
The signal path (SP) consists of the inverse Fourier
transform part as discussed in Fig.\ref{Fig4} and an array  of
mega-pixel camera sensors at the end to sample the intensity profile by
individual 1cm x 1cm cells as discussed with Fig.\ref{Fig5}.
In the SP the probe laser pulses are injected with the polarization tilted by
45 deg with respect to that of the target laser pulses.
In front of the sensors, two polarizers (P) 
selecting photons with the orthogonal combination of polarizations 
so that the birefringence can be measured in a shot, 
which allows the statistical integration of the
measurement over many shots by minimizing the systematic error
due to shot-by-shot fluctuations of the probe pulse energy.
After the implementation of the holographic plate (HP) which produces
the offset phase contrast in the laser interaction zone, we introduce a beam
splitter (BS2) followed by the identical image transferring system as that
in the SP. We refer to this leg as the calibration path (CP).
We classify the origins of local phase fluctuations into the static component
by the optical elements in the paths and the pulse-by-pulse component such as
wavefront fluctuations included in the probe pulse coming from the upstream
laser system. Since the repetition rate of the target laser is limited,
we may inject a single-mode CW laser with the same wavelength as
the dominant part of the probe pulse spectrum into both the SP and the CP,
while the target laser pulses are not injected
(Wavefront aberrations resulting from reflection by $BS1$ are predetermined).
Because a weak CW laser may be realized as a perfect Gaussian beam 
running in the TEM${}_{00}$ mode at a single longitudinal mode, 
we expect to be able to accurately determine
the static phase component as the average value
by using huge photon statistics accumulated
over a long time period for an experiment 
while probe pulses are not injected.
For the pulse-by-pulse component we use the intensity profile observed
at the end of the CP to reconstruct a set of local phases caused by wavefront
fluctuations included in the probe pulse on the pulse-by-pulse basis.

The measurable four types of local phase sets are denoted as
$\phi^{CW}_{SP}$,
$\phi^{CW}_{CP}$,
$\phi^{PLS}_{SP}$, and
$\phi^{PLS}_{CP}$
where superscripts specify cases of CW and pulse laser injections,
respectively, and subscripts refer to the different paths the beams take.
In the following, all phase sets are interpreted
as those defined on the focal plane, even if the local phases are 
actually embedded in different propagation points.
The two phase sets in the SP are expressed by phases $\varphi$'s with
subscripts corresponding to the names of the optical elements along the path
in Fig.\ref{Fig7} as follows:
\beqa\label{eq_CWSP}
\phi^{CW}_{SP} = 
\varphi_{HP} + \varphi_{BS2} + \varphi_{PM1_{SP}}  + \varphi_{PM2_{SP}}
+  \varphi_{P_{SP}},\quad
\eeqa
and
\beqa\label{eq_PLSSP}
\phi^{PLS}_{SP} = \varphi_{PLS} + \phi^{CW}_{SP},
\eeqa
where 
$BS1$ should be removed when the SP is active, and
$\varphi_{PLS}$ is the pure phase set caused by only the
pulse-by-pulse component which is not correctable by the CW laser.
The two phase sets in the CP are expressed as well
\beqa\label{eq_CWCP}
\phi^{CW}_{CP} = \varphi_{HP} + \varphi_{BS2} +
\varphi_{PM1_{CP}}  + \varphi_{PM2_{CP}} + \varphi_{P_{CP}}, \quad
\eeqa
and
\beqa\label{eq_PLSCP}
\phi^{PLS}_{CP} = \varphi_{PLS} + \phi^{CW}_{CP}.
\eeqa
Combining Eq.(\ref{eq_PLSSP}) and (\ref{eq_PLSCP}), we can restore
the offset phase for the probe pulse injection in the SP, $\phi^{PLS}_{SP}$
by the other measured sets of phases as
\beqa\label{eq_SP}
\phi^{PLS}_{SP} = \phi^{PLS}_{CP} - \phi^{CW}_{CP} + \phi^{CW}_{SP}.
\eeqa
This implies that $\phi^{PLS}_{SP}$ can be restored by other
measurable quantities, which is a necessary condition to allow
the correction within the same probe pulse injection in the SP
in the presence of the laser-induced vacuum phase shift.
We finally describe the entire phase set in the focal plane in the SP when
a target laser pulse exists as
\beqa\label{eq_phiall}
\phi = \phi^{PLS}_{SP} + \delta_{vac},
\eeqa
where $\delta_{vac} = \kappa\varphi_t$ as parametrized 
in Eq.(\ref{eq_kappaphi}).
By substituting Eq.(\ref{eq_phiall}) into Eq.(\ref{eq_chi2}),
we can, in principle, 
determine $\kappa$ for the physical template based on the target
laser profile $\varphi_t$.

The template analysis discussed in Sect.~\ref{subsecA} can also be applied to
determine the individual set of phases in the right hand side 
of Eq.(\ref{eq_SP}).
By assigning a square shape to the region $R_i$ in Eq.(\ref{eq_psiW}), 
representing a cell instead of physical template $\varphi_t$ 
in Eq.~(\ref{eq_chi2}), we estimate $\kappa_i$ for each $R_i(X)$.
The number of photons at a point $W_i$ in the image plane contains the
convoluted phase information of the amplitude from all points
in the transverse plane of the probe $N_X$ as seen from Eq.~(\ref{eq_psiW}).
Therefore, as long as the number of sampling points in the image plane $N_W$
is larger than that in the transverse probe profile $N_X$,
we can, in principle, determine a phase set
from Eq.~(\ref{eq_chi2}) by scanning $\kappa_i$
over the expected dynamic range of the phase variation.
The achievable resolution of the phase reconstruction
depends on the scanning step on $\kappa_i$ in the $\chi^2$-test.
As discussed with Fig.\ref{Fig5} the phase-contrast Fourier imaging achieves 
at least the sensitivity of $\sim 10^{-10}$ for the physical phase
shift by sampling the side band of the intensity distribution 
on the image plane.
Therefore, we can introduce the same resolution step to determine $\kappa_i$.
We may measure the initial coarse phase sets {\it a priori} 
by a commercially available wavefront sensor.
From the phase measurement we can extract the set of phases 
at the focal plane by performing Fourier transform from the image plane
back to the focal plane.
Starting from this initial phase set at the focal plane, we perform
the $\chi^2$-test to determine $\kappa_i$ more accurately
by comparing the computed Fourier image 
to the measured intensity at the image plane. 
If the resolution of commercially available wavefront sensors is 
limited to $\sim \lambda/100$,
we would need to repeat the two-dimensional inverse Fourier transform from
the focal plane to the image plane more than $10^8$ times for scanning 
$\kappa_i$, in order to reach the same phase resolution as $\sim 10^{-10}$.
Accordingly, a proper computing power is necessary to restore the sets of
the offset phases in Eq.(\ref{eq_SP}) on the pulse-by-pulse basis.

\subsection{Background in the phase-contrast Fourier imaging}
  \label{subsecC}

A background source of the current measurement is
the refractive index shift due to the plasma creation
from the residual gas along the path of the focused target laser pulse.
The refractive index of the static plasma in the limit of negligible collisions
between charged particles is expressed as
%
\begin{eqnarray}\label{EqNplasma}
N = \sqrt{1-\frac{\omega_p{}^2}{\gamma \omega_0{}^2}},
\end{eqnarray}
%
where
      $\omega_0$ is the angular frequency of the target laser,
      $\omega_p$ is the plasma angular frequency defined as
      $\sqrt{4\pi e^2 n_e / m_e}$ and
      $\gamma$ is the relativistic Lorentz factor
      given as $\sqrt{1+a_0^2}$ with
      $a_0=0.85\times10^{-9}\lambda[\mu m]\sqrt{I_0[W/cm^2]}$.
In the low-pressure limit of the residual gas, the amount of refractive
index shift $\Delta N \equiv N-1$ is expressed as 
$\omega_p{}^2 / 2\gamma \omega_0{}^2$.
Although the refractive index in the plasma becomes smaller than that of
the peripheral area with neutral atoms, the inverted phase
contrast of the phase shift inside the probe pulse still maintains a
rectangular shape along the trajectory of the target laser.
Therefore, it should produce the characteristic diffraction
pattern at similar locations to the nonlinear QED case
as expected from the Babinet's principle,
which requires that the diffraction pattern from an
opaque slit plus the inverted slit of the same size and shape form
an amplitude distribution identical to that of the incident wave
as we discussed in section~\ref{sec2}.
In order to reduce this effect, we need to reduce the electron density
$n_e$ in the residual gas. If we take $\gamma \sim 1$ as the upper limit
of the $\Delta N$ estimate, the refractive index shift
$\sim 10^{-11}$ due to the nonlinear QED effect for a reference
energy density $\sim 1$J/$\mu$m${}^3$, corresponding to a residual gas 
pressure of $\sim 10^{-5}$~Pa. 
The collisional frequency due to interactions between electrons
and ions is expected to be $10^8-10^9$s${}^{-1}$ at the critical electron
density $n_{cr}[cm^{-3}]=1.12\times 10^{21}/\lambda^2[\mu m]$, where $\omega_p$
equals $\omega_0$. For a duration time of $\sim$fs of the target laser pulse,
the inverse bremsstrahlung radiation due to collisional processes
in the residual gas is negligible at $\sim 10^{-5}$~Pa.

Plasma formation is also caused by the probe pulse along its waist over a
distance of $\sim 1$~mm. At a pressure of $\sim 10^{-9}$~Pa the associated
plasma induced phase shift is one order of magnitude smaller than 
that due to QED. Hence the pressure of the residual gas in the interaction 
chamber has to be kept at this level.

We note that the actual processes will be more dynamical,
due to the pondermotive force executed by the high-intensity laser field.
In such a case the refractive index shift based on static plasma
gives only the upper bound on the amount of the local refractive 
and phase shift.

\section{Potential effects beyond QED}\label{sec4}
In the previous sections we discussed the design of the
phase-contrast Fourier imaging by aiming at probing the vacuum 
birefringence through the QED effect, namely, the
electron-positron loop to which photons couple. However,
if the quark mass in vacuum is of the same order as the electron mass,
we should expect that quarks also contribute to the vacuum birefringence
by replacing the electron-positron loop with the quark-antiquark loop.
Whether this effect has a sizable contribution or not
is, however, difficult to quantify with presently existing field theoretical
approaches, because of the strong coupling of quantum-chromodynamics
(QCD) in vacuum, where the coupling is too large to allow for 
a perturbative treatment.
Moreover, bare quark masses not confined in hadrons are not precisely known.
In addition to calculations based on the QCD field theory~\cite{Rafelski1,Rafelski2},
there is another possibility that the duality between string theory with 
higher dimensions and field theory in 3+1 dimensions (holography)
~\cite{Maldacena,MaldacenaReport} could be directly applicable
to this birefringence problem~\cite{Zayakin}.
The QCD and holographic approaches may give different predictions
for the balance of coefficients between the two terms of the Euler-Heisenberg
Lagrangian.
Therefore, 
we may be able to pin down such theoretical issues
by accumulating statistics more than a single shot and also
expecting further increase of the laser intensity in the future.

Moreover,
we note that because the photon-photon scattering cross section 
of QED interaction in the perturbative regime
is so small, $10^{-42}$~b at optical frequency (see \cite{KN,DT}),
we experience little 'noise', providing
a pristine experimental environment to search for something beyond QED.
Suppose then the detected dispersion and birefringence quantitatively deviate
from the expectation of QED, including potential QCD corrections.
This should indicate that undiscovered fields
may be mediating photons beyond QED and QCD.
Scalar and pseudoscalar types of fields in vacuum
may contribute via the first and second
products in the brackets of Eq.~(\ref{eq_EHL}), respectively.
They may be candidates of cold dark matter, if the coupling to photons
and the mass are reasonably small~\cite{PDG}. 
Therefore, the measurement of the absolute value of the phase shift
depending on the polarization combinations and the comparison to
the expectations from nonlinear QED including potential QCD corrections
may be a general test of unknown nature in vacuum.

\section{Conclusion}\label{sec5}
We suggest an approach to probe the vacuum birefringence
under the influence of intense lasers.
The phase-contrast Fourier imaging technique
can provide a sensitive method to measure the absolute phase shift of 
light crossing intense laser fields.
With this method nonlinear QED effects of the Euler-Heisenberg Lagrangian
may be detected requiring no more than lasers of the hundred PW-class.
The method provides a window for scoping the vacuum via the dynamics of
the electron mass scale and possibly the lightest quark mass.
Such a detection has never been made to date, and it heralds the research in
the physics of the vacuum with a high-field approach.
Given the high-intense optical lasers
available in the ELI project~\cite{ELI} in the near future, the realization of 
this suggestion may become an exciting challenge for future experiments
exploring vacuum physics.

\vspace{1cm}

{\bf Acknowledgment}\\
This research has been supported by the DFG Cluster of Excellence MAP
(Munich-Center for Advanced Photonics).
K. Homma appreciates the support by the Grant-in-Aid for Scientific Research 
no.21654035 from MEXT of Japan. 
T. Tajima is Blaise Pascal Chair Laureate 
at the \'Ecole Normale Sup\'erieure.
We thank H. Gies and S. Sakabe for their advices and P. Thirolf for
his careful reading of our manuscript.

\end{document}